\documentclass[aip,reprint,preprintnumbers,noshowpacs,noshowkeys]{revtex4-1}
\usepackage[T1]{fontenc}
\usepackage{times}
\usepackage{graphicx,color}
\usepackage{amsfonts,amsmath,amssymb,amsbsy}
\usepackage[colorlinks=true, linkcolor=blue, citecolor=blue, urlcolor=blue]{hyperref}

\let\mathbf=\boldsymbol

\def\blue#1{\textcolor{blue}{#1}}
\def\blue#1{\textcolor{black}{#1}}
\def\emph#1{\textcolor{blue}{#1}}
\begin{document}

\title{A spiking neuron constructed by the skyrmion-based spin torque nano-oscillator}

\author{Xue Liang}
\affiliation{College of Physics and Electronic Engineering, Sichuan Normal University, Chengdu 610068, China}
\affiliation{School of Science and Engineering, The Chinese University of Hong Kong, Shenzhen, Guangdong 518172, China}

\author{Xichao Zhang}
\affiliation{School of Science and Engineering, The Chinese University of Hong Kong, Shenzhen, Guangdong 518172, China}

\author{Jing Xia}
\affiliation{School of Science and Engineering, The Chinese University of Hong Kong, Shenzhen, Guangdong 518172, China}

\author{Motohiko Ezawa}
\affiliation{Department of Applied Physics, The University of Tokyo, 7-3-1 Hongo, Tokyo 113-8656, Japan}

\author{Yuelei Zhao}
\affiliation{School of Science and Engineering, The Chinese University of Hong Kong, Shenzhen, Guangdong 518172, China}

\author{\\ Guoping Zhao}
\email[E-mail:~]{zhaogp@uestc.edu.cn}
\affiliation{College of Physics and Electronic Engineering, Sichuan Normal University, Chengdu 610068, China}

\author{Yan Zhou}
\email[E-mail:~]{zhouyan@cuhk.edu.cn}
\affiliation{School of Science and Engineering, The Chinese University of Hong Kong, Shenzhen, Guangdong 518172, China}

\begin{abstract}
Magnetic skyrmions are particle-like topological spin configurations, which can carry binary information and thus are promising building blocks for future spintronic devices. In this work, we investigate the relationship between the skyrmion dynamics and the characteristics of injected current in a skyrmion-based spin torque nano-oscillator, where the excitation source is introduced from a point nano-contact at the center of the nanodisk. It is found that the skyrmion will move away from the center of the nanodisk if it is driven by a spin-polarized current; however, it will return to the initial position in the absence of stimulus. Therefore, we propose a skyrmion-based artificial spiking neuron, which can effectively implement the leaky-integrate-fire operation. We study the feasibility of the skyrmion-based spiking neuron by using micromagnetic simulations. Our results may provide useful guidelines for building future magnetic neural networks with ultra-high density and ultra-low energy consumption.
\end{abstract}

\date{24 March 2020}

\preprint{\textit{Appl. Phys. Lett.} \textbf{116}, 122402 (2020); DOI:~\href{https://doi.org/10.1063/5.0001557}{10.1063/5.0001557}}

\maketitle


The neural network, biologically-inspired computing model, is a recent hot topic in the field of artificial intelligence, which can be used for pattern recognition, data classification and prediction tasks.~\cite{LeCun_NATURE2015,Prezioso_NATURE2015,Kuzum_NANOTECHNOL2013} The human brain containing nearly $86$ billion neurons and each of which has thousands of synapses connected to other neurons, can quickly complete many cognitive tasks even from an imperfect set of input information, with ultra-low energy consumption.~\cite{Grollier_IEEE2016,Sengupta_APR2017,Sengupta_APE2018} However, the current neuron network based on conventional complementary metal-oxides-semiconductor (CMOS) hardware~\cite{Wu_IEEE2015,Chu_IEEE2015} is a simple abstract model of the human brain. Consequently, there are two significant bottlenecks. Firstly, similar to the biological brain, the neural network consists of a large number of artificial neurons interconnected by synapses to handle complex relationships among data, so that the scale of networks is limited.~\cite{Grollier_IEEE2016,Sengupta_APR2017,Sengupta_APE2018,Indiveri_IEEE2015} Secondly, such tremendous hardware resources require a large power supply, which is orders of magnitude higher than that of human brain.~\cite{Sengupta_APR2017,Sengupta_APE2018} Therefore, in order to optimize the artificial neural network, it is crucial to find potential device structures for neurons and synapses to shrink the scale of the network as well as to reduce the energy consumption.

On the other hand, magnetic skyrmions are topologically non-trivial objects,~\cite{Rossler_NATURE2006,Nagaosa_NNANOTECHNOL2013,Finocchio_JPD2016,Wiesendanger_MATER2016,Kang_IEEE2016,Fert_MATER2017,Wanjun_PHYS2017,Everschor-Sitte_JAP2018,Xichao_JPC2019} which are experimentally discovered in chiral magnet MnSi~\cite{Muhlbauer_SCIENCE2009} ten years ago. Since then, magnetic skyrmions have aroused a lot of interest due to their attractive physics.~\cite{Zhou_NCR2019} Compared with the traditional counterparts, such as magnetic bubbles and domain walls, skyrmions have nanoscale size, stable and rigid structure, as well as ultra-low depinning current density.~\cite{Nagaosa_NNANOTECHNOL2013,Zhou_NCR2019,Jonietz_SCIENCE2010,Sampaio_NNANOTECNOL2013,Iwasaki_NANOTECHNOL2013,Xichao_SCI7643,Woo_NMATER2016} With these unique features, skyrmions could be used as information carriers for next-generation nanoscale magnetic storage and logic devices.~\cite{Tomasello_SCI2014,Xichao_SCI9400} Most recently, skyrmions are also observed in different material systems at room temperature, such as asymmetric ferromagnetic multilayer stacks,~\cite{Woo_NMATER2016,Wanjun_SCIENCE2015,MoreauLuchaire_NNANOTECHNOL2016,Boulle_NNANOTECHNOL2016,Guoqiang_NANOLETTERS2016,Soumyanarayanan_MATER2017} frustrated kagome magnets~\cite{Hou_NANOLETTERS2018} and ferroelectric materials,~\cite{Das_NATURE2019} which lays the foundation for their future applications in spintronic devices.

\begin{figure}[t]
\centerline{\includegraphics[width=0.50\textwidth]{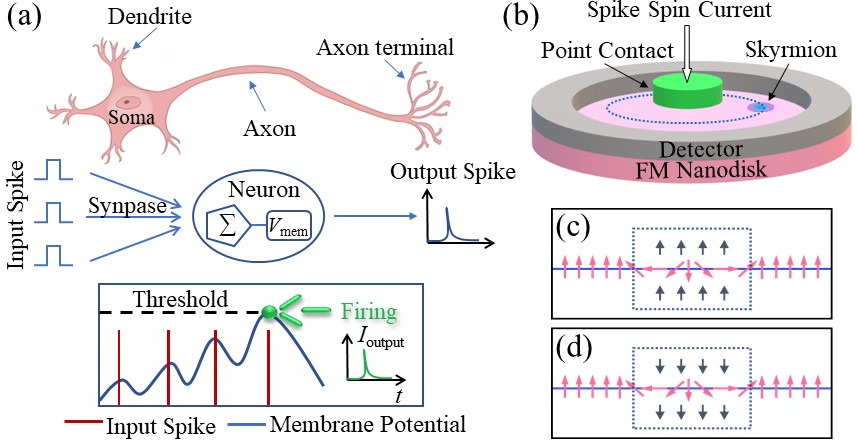}}
\caption{%
(a) Demonstration of the structure and function of a biological neuron. Here, $\Sigma$ means that the soma of the neuron integrates all the excitatory and inhibitory input signals.
(b) Schematic of the proposed spiking neuron. 
(c-d) Illustration of the directions of spin-current polarization and local magnetic moments.
}
\label{FIG1}
\end{figure}

Taking all advantages of skyrmions into account, the challenges of current neural network mentioned above may be resolved by using skyrmions in artificial neurons and synapses. For examples, there are several theoretical reports on this topic.~\cite{Huang_NANOTECHNOL2017,Sai_NANOTECHNOL2017,Pinna_PRAPPLIED2018,Chen_NANOSCALE2018,Chen_IEEE2018,He_DATE2017,Azam_JAP2018,Prychynenko_PRAPPLIED2018}
In this work, we proposed a design of artificial neuron based on magnetic skyrmions, which can be used in the spiking neural networks (SNNs).~\cite{Maass_1997} Similar to the biological neuron [see Fig.~\ref{FIG1}(a)], the proposed neuron integrates frequently input stimuli through a membrane potential (i.e., the skyrmion position) and fires an output spike once the potential reaches a certain threshold. Such a neuron is also referred to as ``leaky-integrate-fire'' (LIF) spiking neuron in neural networks.~\cite{Stoliar_2017,Burkitt_2006,Brette_2005,Burkitt_22006,Wesley_2019}

As illustrated in Fig.~\ref{FIG1}(b), a nanoscale point-contact device is employed to realize the function of the spiking neuron, which includes a skyrmion-based spin torque nano-oscillator (STNO),~\cite{Zhang_NJP2015} a ring-shape detection electrode, and peripheral circuits.The center point nano-contact electrode is used to create and drive a skyrmion by injecting spin-polarized currents with different amplitudes, and the detection electrode is used to detect the voltage signal induced by the existence of skyrmions due to the magneto-resistive (MR) effect. The ferromagnetic nanodisk is attached to a heavy metal layer with strong spin-orbit coupling,~\cite{Finocchio_JPD2016,Kang_IEEE2016} which can provide an interface-induced Dzyaloshinskii-Moriya Interaction (DMI)~\cite{Dzyaloshinsky_1958,Moriya_1960} to stabilize the skyrmion. The energy density related to DMI is given by~\cite{Rohart_PRB2013} 
\begin{equation}
\begin{split}
\label{eq:EDMI}
 \varepsilon_{\text{DMI}}=D[m_{z}(\boldsymbol{\nabla}\cdot\boldsymbol{m})-(\boldsymbol{m}\cdot\boldsymbol{\nabla})m_{z}], 
\end{split}
\end{equation}
where $D$ is the DMI constant, $\boldsymbol{m}=\boldsymbol{M}/M_{\text{S}}$ is the local magnetization reduced by the saturation magnetization $M_{\text{S}}$, and $m_z$ is the out-of-plane component of the reduced magnetization. The total energy of the system is written as
\begin{equation}
\begin{split}
\label{eq:E}
\varepsilon_{\text{total}}=&\int \{A[({\nabla}m_x)^2+({\nabla}m_y)^2+({\nabla}m_z)^2] \\
&-K{m_{z}}^2+\varepsilon_{\text{DMI}}+\varepsilon_d\,\} {d}x{d}y{d}z,
\end{split}
\end{equation}
where $A$ and $K$ are the Heisenberg exchange stiffness and perpendicular magnetic anisotropy (PMA) constant, respectively. The last term $\varepsilon_{\text{d}}$ denotes the demagnetization energy density.

To simulate the spin dynamics in the proposed device, we use the public domain micromagnetics program Object Oriented Micromagnetic Framework (OOMMF)~\cite{OOMMF} with extension DMI module, which solves the Landau-Lifshitz-Gilbert (LLG) equation
\begin{equation}
\begin{split}
\label{eq:LLG} 
\frac{d\boldsymbol{m}}{dt}=-\gamma\boldsymbol{m}\times\boldsymbol{H}_{\text{eff}}+\alpha\boldsymbol{m}\times{\frac{d\boldsymbol{m}}{dt}}+\boldsymbol\tau_{\text{T}},
\end{split}
\end{equation}
where $\gamma$ is the gyromagnetic ratio, $\alpha$ is the Gilbert damping coefficient, $\boldsymbol{H}_{\text{eff}}=-\mu_{0}^{-1}\frac{\partial \varepsilon}{\partial \boldsymbol{m}}$ is the effective field related to various micromagnetic energies. The last term on the right-handed side of Eq.~(\ref{eq:LLG}) stands for the spin torque induced by the spin-polarized current,~\cite{Sampaio_NNANOTECNOL2013} which can be written as $\boldsymbol\tau_{\text{T}}=u\boldsymbol{m}\times(\boldsymbol{m}\times\boldsymbol{m}_{\text{p}})$. Here, $u=\frac{{\gamma\hbar}jP}{2e\mu_{0}M_{\text{S}}{t_z}}$, $\hbar$ is the Plank constant, $P=0.4$  is the assumed spin polarization value, $j$ is the spin-polarized current density, $\boldsymbol{m}_{\text{p}}$ is the spin polarization direction along the +z-direction, $e$ is the electron charge, $\mu_{0}$ is the vacuum permeability and $t_{z}$ is the thickness of the ferromagnetic nanodisk. By numerically solving Eq.~(\ref{eq:LLG}), one can obtain the evolution of local spin with time, and determine the dynamics of a skyrmion.

In our simulations, the magnetic material parameters are adopted from Ref.~\onlinecite{Sampaio_NNANOTECNOL2013}: the saturation magnetization $M_{\text{S}}=0.58$ MA/m, exchange stiffness $A=15$ pJ/m, DMI constant $D=3.0$ mJ/m$^2$, PMA constant $K=0.8$ MJ/m$^3$, the damping coefficient $\alpha=0.1$. We assume that the nanodisk has a fixed diameter of $100$ nm and a thickness of $0.4$ nm. The mesh size of $1 \times 1 \times 0.4$ nm$^3$ is used to discretize the model.

\begin{figure}[t]
\centerline{\includegraphics[width=0.45\textwidth]{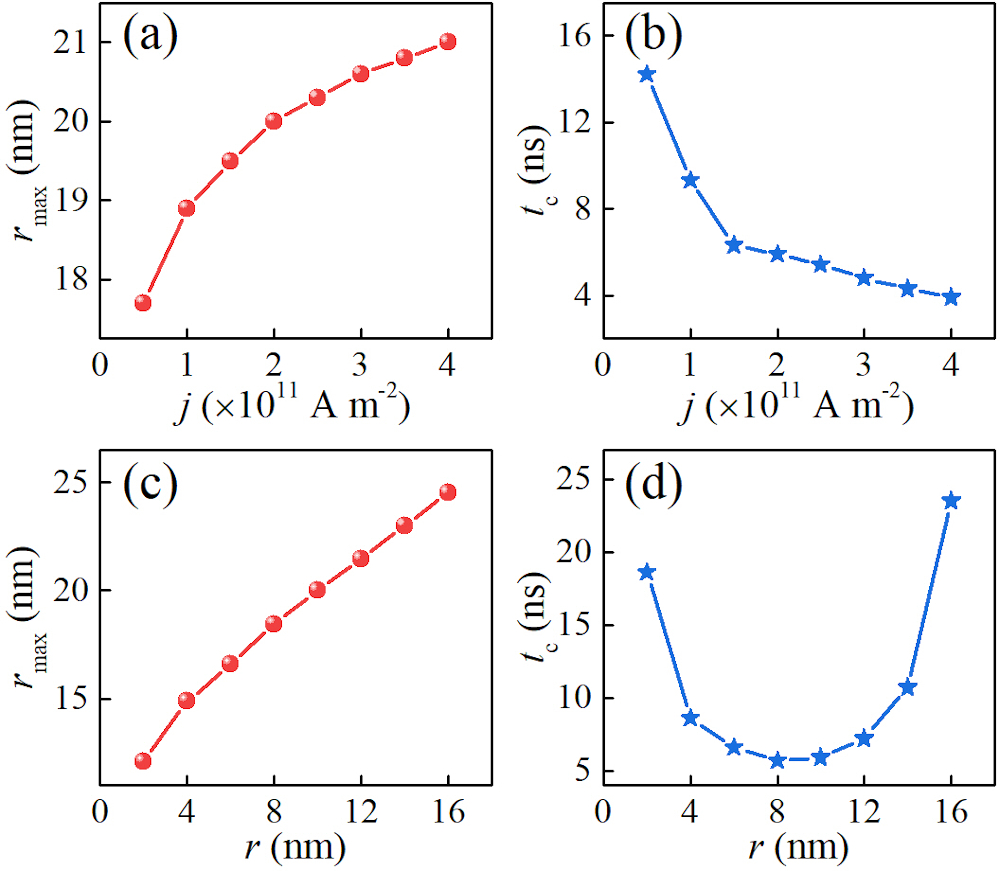}}
\caption{%
The maximum motion radius of skyrmion $r_{\text{max}}$ and the critical time $t_c$ that skyrmion starts to do a stable circle motion as a function of (a-b) the current density $j$ and (c-d) the injection radius $r$.
}
\label{FIG2}
\end{figure}

As reported in Ref.~\onlinecite{Zhang_NJP2015}, the skyrmion at the center of the nanodisk will make a spiral motion and finally reach a persistent oscillation if a spin-polarization current is injected perpendicularly from the nano-point-contact electrode. However, we find that such a dynamic behavior occurs when the directions of the polarization $\boldsymbol{m}_{\text{p}}$ and the core spins of a skyrmion are opposite. If the polarization direction and skyrmion core spins are aligned in the same direction, the skyrmion will remain stationary.

Figure~\ref{FIG1}(c) and \ref{FIG1}(d) schematically illustrates the distribution of magnetic spins along the radial direction of the nanodisk when the skyrmion is at the nanodisk center. From the phenomenological point of view, if the direction of the local magnetic moment is opposite to the spin-current polarization, it will suffer a torque and tend to be parallel to the polarization.~\cite{Finocchio_JPD2016,Tomasello_SCI2014} This is the reason why the skyrmion size reduces first in the simulation. However, considering the existence of other interactions in the given system, the skyrmion will not continue to shrink, unless the current density is large enough. Therefore, for a mediate current density, the skyrmion will deviate from the center area of the nanodisk to reduce the energy of the entire system containing conductive electrons (i.e., the skyrmion is subject to an effective centrifugal force).

As the skyrmion gradually moves away from the nanodisk center, the repulsive force from the nanodisk boundary increases and eventually balances with the centrifugal force induced by driving current.Therefore, the skyrmion shows a perfect circle motion. On the contrary, if the directions of the polarization and the core spins of skyrmion are the same, the skyrmion will expand and cannot be driven into motion since the energy of the system is lower when the skyrmion is at the center of nanodisk. Therefore, the spin polarization direction $\boldsymbol{m}_{\text{p}}$ is set as $+z$, and the core spins of skyrmion point down in our following simulation.

We further investigate the effect of two other parameters of the driving current, namely, the driving current density $j$ and the injection area radius $r$, on the skyrmion motion behaviors in the nanodisk.
Here, we mainly focus on the critical time $t_c$ and the corresponding maximum motion radius $r_{\text{max}}$ when the skyrmion starts to do a steady circular motion.
As shown in Figure~\ref{FIG2}(a) and~\ref{FIG2}(b), when the injection radius is fixed, the skyrmion will reach the dynamic equilibrium state faster for a lager current density. Also, the maximum radius $r_{\text{max}}$ increases as the current density increases. However, the speed of growth gradually decreases, which is attributed to the boundary restriction of the nanodisk. In fact, the skyrmion is confined in the nanodisk, which limits the increase of motion radius, i.e., the skyrmion motion is a result of the competition between the effective centrifugal force and the repulsive force from the boundary. When the current density is small, the maximum radius $r_{\text{max}}$ mainly depends on the effective driving force. For a lager driving current density, the boundary-induced force limits the maximum radius $r_{\text{max}}$.
%
\begin{figure}[t]
\centerline{\includegraphics[width=0.5\textwidth]{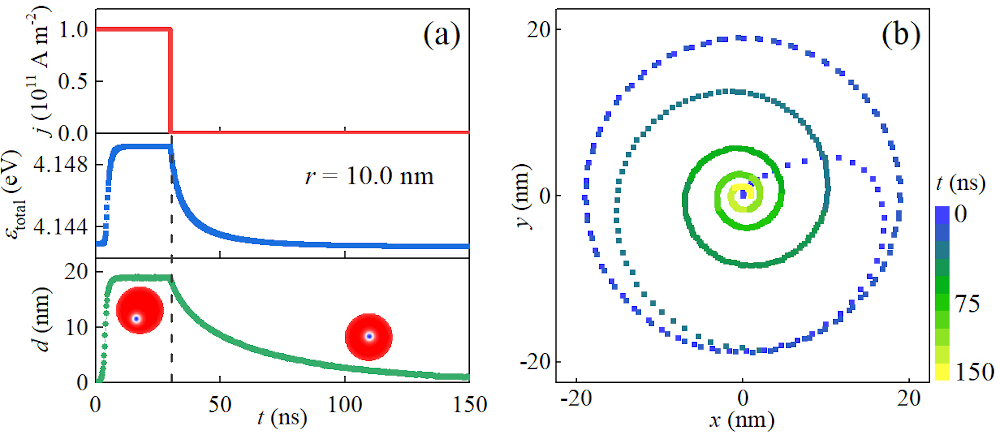}}
\caption{%
(a) The current density $j$, total energy $\varepsilon_{\text{total}}$ and the distance $d$ between the skyrmion and the center of nanodisk as a function of the time, where the injection radius $r$ is fixed at $10$ nm. Insets are the snapshots of simulations, where the out-of-plane component of magnetization is color coded: red is out-of-plane, blue is in into plane, white is in-plane. 
(b) Micromagnetic simulation of the trajectory of skyrmion motion in a nanodisk.
}
\label{FIG3}
\end{figure}

One the other hand, Fig.~\ref{FIG2}(c) and~\ref{FIG2}(d) show the effect of the injection radius $r$ on the skyrmion dynamics, where the driving current density is fixed at $2.0\times10^{11}$ A/m$^2$. It is found that the maximum radius $r_{\text{max}}$ is proportional to the injection radius. However, there is a local minimum value of the threshold time $t_c$ when $r\sim{r_{\text{sk}}}$ (where the $r_{\text{sk}}$ is the radius of the skyrmion approximately $8.0$ nm in our work). When $r$ < $r_{\text{sk}}$, the critical time $t_c$ decreases as the injection radius increases. On the contrary, for $r$ > $r_{\text{sk}}$, the critical time $t_c$ is proportional to the injection radius. The reason could be that the driving force is opposite to the centrifugal force when the spin current flows through the regions outside the skyrmion core (i.e., the area of $m_z$ < $0$), which prevents the skyrmion moving away from the center of the circle. As a result, the total effective centrifugal force will be reduced and the critical time will lengthen.

Considering a simple time-varying current given in Fig.~\ref{FIG3}(a), the trajectory of skyrmion is demonstrated in Fig~\ref{FIG3}(b). The skyrmion initially deviate from the center of nanodisk due to the effective centrifugal force induced by driving current, as mentioned above. However, if the current is suddenly removed, the skyrmion will make a spiral motion and gradually return to the initial position due to the repulsive force from boundary. From the perspective of energy, the total energy of the system will decrease when the skyrmion approaches to the center of nanodisk, as shown in Fig.~\ref{FIG3}(a). Therefore, the skyrmion will relax to the center of nanodisk. Fig~\ref{FIG3}(a) shows the distance $d$ between the center of nanodisk and skyrmion versus the time.
Also, the relationship between the energy and the skyrmion position is given in Supplemental Material.

\begin{figure}[t]
\centerline{\includegraphics[width=0.48\textwidth]{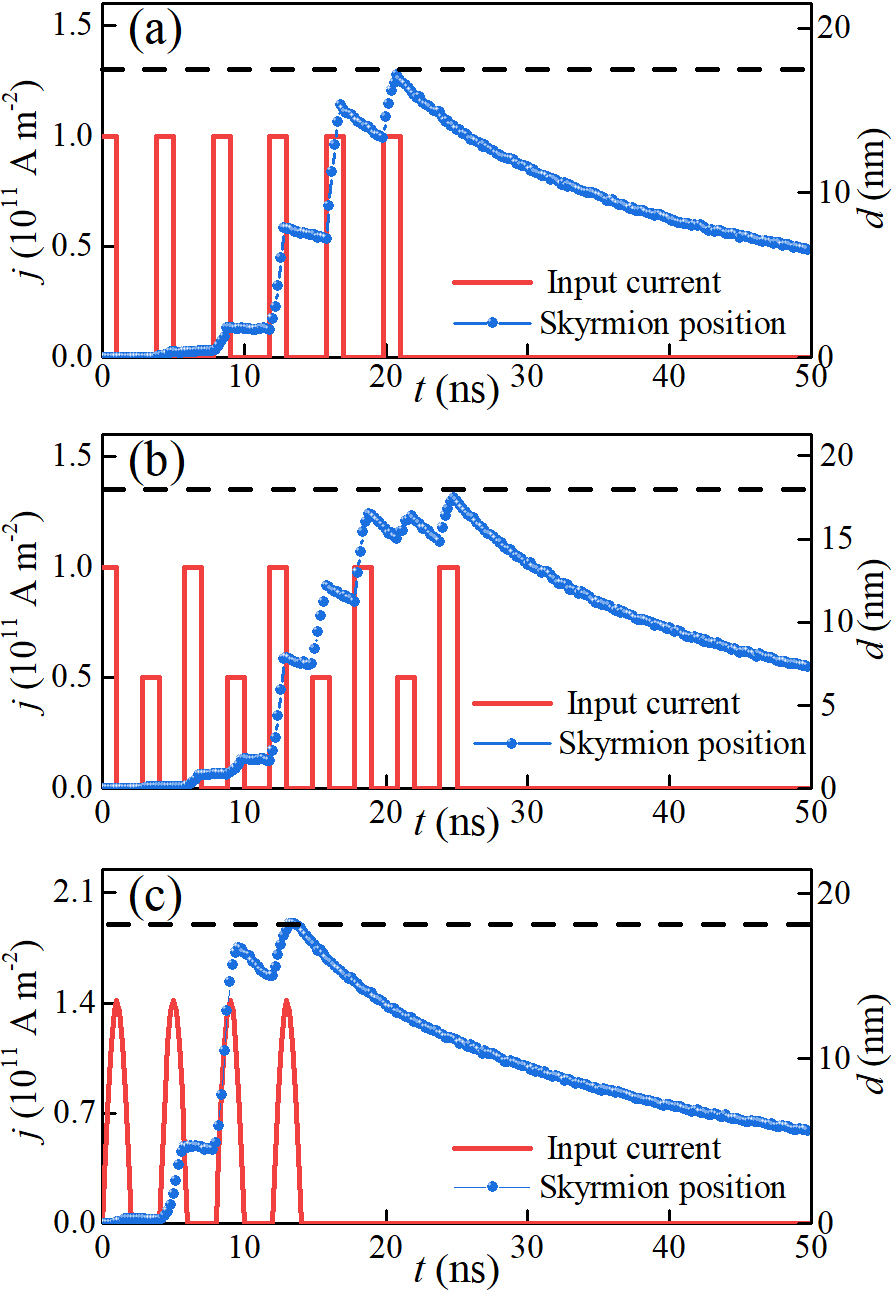}}
\caption{%
Skyrmion motion with three different spike current signals.
(a) The homogeneous square wave with frequency of $0.33$ GHz.
(b) The inhomogeneous square wave with frequency of $0.5$ GHz.
(c) The sinusoidal wave with period of $2.0$ ns.
}
\label{FIG4}
\end{figure}

Such a back and forth motion of skyrmion driven by the current can be used to mimic the LIF spiking neuron operation, which is inspired by the behavior of a biological neuron.
In the biological nervous system, one neuron has many dendrites, which are connected to the axon terminals of other neurons through synapses to realize the transmission and processing of information. The input signals from pre-neuron are converted into neurotransmitters and then act on the post-neuron. As the input signals from dendrites arrive, the potential difference (i.e., the membrane potential) between the inside and the outside of the neuron increases. It should be noted that the membrane potential will decrease if no signal arrives at some intervals, which corresponds to the leakage of potential. As shown in Fig.~\ref{FIG1}(a), when the potential accumulates to a threshold voltage, the soma generates an action potential (also referred to as a neuron spike), which is then transmitted to other neurons through synapses.
If we consider this distance $d$ between the center of nanodisk and skyrmion as the membrane potential of the artificial neuron, the proposed device can well perform the function of LIF spiking neuron. The detector is initially placed at a distance of $d_c$ from the center of nanodisk, which must be smaller than the maximum motion radius of skyrmion. With continuous injection of the spike current, skyrmion will arrive at the threshold distance $d_c$ and be detected by the detector, at which the neuron fires a spiking signal and resets.
\blue{It should be noted that one can increase the current density, the injection radius or the diameter of the nanodisk to increase the threshold distance, which is about $17$ nm in our simulation.}
On the other hand, if we apply a current whose direction is opposite to the driving current at the nano point contact, the skyrmion can quickly return to the center of the nanodisk.~\cite{Sai_NANOTECHNOL2017,Chen_NANOSCALE2018}
\blue{Once the skyrmion is destroyed, a new skyrmion can be created under the nano-contact by injecting a different spin polarized current.}

\begin{figure}[t]
\centerline{\includegraphics[width=0.48\textwidth]{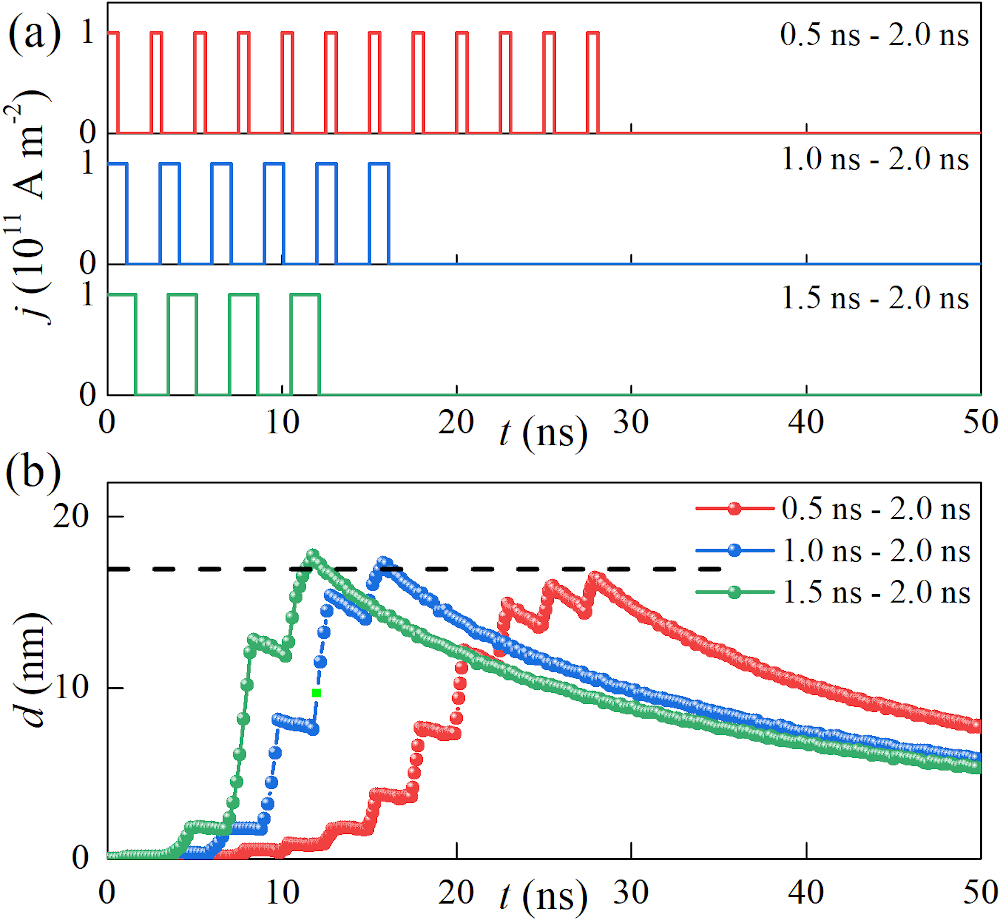}}
\caption{%
Skyrmion motion under three spike current with different action times $t_1$ of each current pulse. 
(a) Three specific input spike current with the same time interval between signals $t_{2} = 2.0$ ns.
(b) Location of the skyrmion as a function of time.
}
\label{FIG5}
\end{figure}

Considering there are various kinds of input signals in actual neurons, we simulate the response of the skyrmion to different types of input currents to verify the feasibility of our design.
Fig.~\ref{FIG4} shows the dynamic behaviors of skyrmion under three different spike currents which are periodic homogeneous square wave, inhomogeneous square wave and sinusoidal wave.
\blue{Note that selected top-view snapshots of simulated magnetization configurations are given in Supplemental Material.}
It can be seen that the proposed neuron based on skyrmions can well implement the leaky-integrate-fire operation no matter what type of signal is input. We also discuss the response of skyrmion to different intervals of the spike input current.
As shown in Fig.~\ref{FIG5}, if the action time of each current pulse is changed, the amount of ``integration'' will increase for a pulse with a longer duration. Consequently, the number of required signals decreases, and the neuron fires an output signal quickly (see Supplemental Material).
\blue{In addition, the vortex-type STNO~\cite{Locatelli_SCI2015,Khvalkovskiy_PRB2009,Guslienko_NRL2014,Belanovsky_PRB2012,Jiang_APL2019,Ma_JPD2020,Siracusano_PRL2016} has been studied extensively, which may be used as an alternative to the skyrmion-based STNO to implement basic functionality of an artificial spiking neuron. Therefore, we also simulate the artificial spiking neuron based on a vortex-type STNO (see Supplemental Material).}

In particular, the magnetic skyrmion is a rigid spin structure with nanoscale size and good stability,~\cite{Varentcova_arXiv2020} so that this design can effectively improve the space utilization in future spiking neural network.
\blue{On the other hand, several studies have reported that skyrmions can be stabilized at room temperature in the form of either skyrmion lattices or isolated skyrmions.~\cite{Woo_NMATER2016,Wanjun_SCIENCE2015,MoreauLuchaire_NNANOTECHNOL2016,Boulle_NNANOTECHNOL2016,Guoqiang_NANOLETTERS2016,Soumyanarayanan_MATER2017,Chen_APL2015} For example, Boulle et al.~\cite{Boulle_NNANOTECHNOL2016} observed stable room-temperature skyrmions in sputtered ultrathin Pt/Co/MgO nanostructures under zero external magnetic field, which is a direct experimental evidence that skyrmions are promising building block for future room-temperature spintronic devices.}
In addition, micromagnetic simulations reveal that the current density to drive skyrmions is about $10^{11}$ A/m$^2$ and the injection area is a nano point contact with a diameter of $10$ nm, which can reduce the power consumption significantly.

In conclusion, we have investigated the effect of three parameters of input current including the direction of polarization, the current density and the injection radius, on the dynamic behaviors of a skyrmion in a skyrmion-based STNO structure. Inspired by the back and forth motion of skyrmion induced by time-dependent driving currents, we proposed a LIF spiking neuron based on skyrmions, whose functionality and performance are also verified by micromagnetic simulations. Our results are useful for fabricating energy-efficient neurons for constructing future spiking neural network at nanoscale.

\blue{See the supplementary material for more results about the relation between the total energy and skyrmion position, the snapshots of simulated skyrmion motion, the response of the artificial neuron to input signals with different intervals, and the simulation of the spiking neuron based on a vortex-type STNO.}

\vbox{}


X.Z. was supported by the Guangdong Basic and Applied Basic Research Fund (Grant No. 19201910240003361) and the Presidential Postdoctoral Fellowship of The Chinese University of Hong Kong, Shenzhen (CUHKSZ).
M.E. acknowledges the support from the Grants-in-Aid for Scientific Research from JSPS KAKENHI (Grant Nos. JP18H03676, JP17K05490 and JP15H05854) and also the support from CREST, JST (Grant Nos. JPMJCR16F1 and JPMJCR1874).
G.Z. acknowledges the support by the National Natural Science Foundation of China (Grant Nos. 51771127, 51571126 and 51772004) of China, the Scientific Research Fund of Sichuan Provincial Education Department (Grant Nos. 18TD0010 and 16CZ0006).
Y.Z. acknowledges the support by the President's Fund of CUHKSZ, Longgang Key Laboratory of Applied Spintronics, National Natural Science Foundation of China (Grant Nos. 11974298 and 61961136006), Shenzhen Fundamental Research Fund (Grant No. JCYJ20170410171958839), and Shenzhen Peacock Group Plan (Grant No. KQTD20180413181702403).



\end{document}